# Demand Side Management for Homes in Smart Grids

Mohammad Rasoul Narimani
Department of Electrical and Computer Engineering
Missouri University of Science and Technology
Rolla, USA
Mn9t5@mst.edu

*Abstract*—**Electricity usage is a major portion of utility bills and the best place to start lowering them. An effective home energy management approach is introduced to decrease customers' electricity bills by determining the optimal appliance scheduling under hourly pricing based demand response (DR) strategies. The proposed approach specifically addresses consumer comfort through acceptable appliance deferral times, within which their consumption can be shifted from the normal schedules. Electric vehicles (EVs), with vehicle-to-grid (V2G) and vehicle-to-home (V2H) abilities, Energy storage systems (ESS), and Renewable Energy Resource as a distributed generation are considered at the end-user premises to improve the energy management efficiency. A mixed-integer linear programming (MILP) framework is implemented to solve the proposed problem. Numerical results show that the proposed approach can reduce customers' electricity bill without sacrificing their comfort and fully exploit EVs and ESS capacities.**

*Index Terms*--**demand side management, renewable energy sources, electric vehicles, flexible loads, smart grid.**

## I. INTRODUCTION

Demand side management (DSM), a key component of future smart grids, can increase the efficiency of electric grids by modifying consumer's electric demand through various methods such as financial incentives and behavioral change [1]-[4]. DSM focuses on interacting between utility and customers. These interactions include exchange power between electric grid and customer and also modifying demand profile based on electricity prices at different time intervals [5], [6]. Presenting an algorithm for obtaining the best demand management is crucial to fully exploit the DSM capabilities.

Most of the DSM-related work has focused on large customers that can make a significant contribution to grid balance. The power systems literature includes a plethora of studies regarding implementing energy management techniques in larger electric customers [7]. Emerging load management techniques that can control end user's loads facilitates implementing DSM on residential loads [8]. However, different appliances along with energy storage systems (ESSs) and electric vehicles (EVs) in residential customers makes the energy management problem more complicated [9]. Another important point that distinguishes DSM in residential sector from industrial is observing customer's comfort. The best residential DSM approach is one that can manage loads without sacrificing customer's comfort [10], [11].

There is limited research related to home energy management with customers' comfort consideration. A real-time priority-based load scheduling model based on the availability of renewable energy resources and dynamic pricing scheme is presented in [12] to maximize benefits of renewable energy sources and minimize total cost of energy consumption. Consumers' comfort are considered through few constraints based on appliances priority. Priority-based approach cannot fully satisfy customers' comfort and benefit maximization at the same time. Thus, a smarter way for considering customers' comfort is in most of interest which is address in this paper.

The success of any DSM approach strongly depends on its strategy to consider customers' comfort. In this connection, customers' comfort is modeled by defining an acceptable delay times (ADTs) for each appliance. An ADT defines as the maximum time period by which the operation of the associated appliance can be prolonged without disrupting the customers' comfort. Information on the ADT value for different appliances can be obtained by exploring consumers' preferences. Note that the ADT for nondeferable loads is zero. In other words, nondeferable loads cannot be rescheduled from their preferred operation time. Recent advancements in power electronic converters makes the vehicle to grid paradigm a viable option for smart homes [13-16]. The proposed price-based model addresses demand side management of appliances, energy storage system, electric vehicles, photovoltaic power, and two ways energy trading (V2G, V2H, ESS2G and ESS2H) which is unprecedented in energy management literature. Another contribution of this paper is including more precise models of flexible loads whose diverse characteristics were represented as ADTs in home energy management. Also, the proposed method can handle the different load profiles and characteristics by defining different ADTs for different loads. Thus, handling a wider variety of residential, commercial and industrial loads can be easily done with the proposed framework.

## II. PROBLEM FORMULATION

Smart meters play a crucial role in home energy management by processing receiving data from utility, making decision on shifting appliance's load and sending back data to utilities. Smart meters measure households' electricity usage in short increments. A wireless link notifies utility by customers' usage and aware customers by the current electricity price. The latest smart meters are designed to control appliances to reduce customers' electricity bills. The heart of smart meters is a software that process electricity prices (day ahead or real time prices) and load demands to optimize a specific objective function by determining the best combination of loads.

The proposed energy management approach minimizes the electricity cost, equation (1), while satisfies problem's restrictions through multiple equality and inequality constraints. The system loads divide into deferable and nondeferable loads in which deferable loads can be shifted within their ADT limits [17], [18]. Four types of appliances with different ADT including dishwashers, clothes washers, clothes dryers, and heating and ventilation systems are considered as deferable loads. The proposed objective function along with problem's constraints are explained in the next section.

### A. Objective Function

The proposed function for electricity cost includes two different parts. The first part calculates nested electricity cost and the second part is a priority enforcement for EV, ESS and PV to sell their electricity.

$$TC = \sum_t (P_t^{grid} \lambda_t^{buy} \Delta t - P_t^{sold} \lambda_t^{sell} \Delta t) + \sum_t (\varepsilon_1 P_t^{PV,sold} \Delta t + \varepsilon_2 P_t^{ESS,sold} \Delta t + \varepsilon_3 P_t^{EV,sold} \Delta t \quad (1)$$

where $P_t^{grid}$ is withdrawal power from the grid, $\lambda_t^{buy}$ is electricity price at $t^{th}$ time interval, $P_t^{sold}$ is net injected power to the grid including excess power from PV, ESS, and EV ($P_t^{PV,sold}$, $P_t^{EES,sold}$, and $P_t^{EV,sold}$). $\lambda_t^{sell}$ is price of electricity that sell back to the grid. To prevent degrading batteries by frequent charging and discharging, three different coefficients are augmented in objective function to enforce priorities on purchasing power from EV, ESS and PV ($\varepsilon_1 < \varepsilon_2 < \varepsilon_3$).

### B. Power Balance Constraint

The power balance in equation (2), that implies energy conservation principle, must be satisfied to reliably electrify all the loads at home.

$$P_t^{grid} + P_t^{PV,used} + P_t^{EV,used} + P_t^{ESS,used} = PD_t + P_t^{EV,ch} + P_t^{ESS,ch} , \forall t \quad (2)$$

where $P_t^{PV,used}$, $P_t^{ESS,used}$, and $P_t^{EV,used}$ are the withdrawal powers from PV, ESS, and EV to electrify loads at home. $PD_t$ is the total electricity demand at $t^{th}$ time interval. $P_t^{EV,ch}$ and $P_t^{ESS,ch}$ are EV and ESS charging power at $t^{th}$ time interval.

### C. ESS Model

Two different operational modes including ESS2Grid, ESS2Home are considered in the proposed ESS model. Accordingly ESS discharged power can be used to electrify loads at home ($P_t^{EES,used}$) or inject to grid ($P_t^{EES,sold}$). Equation (6) relates ESS's state of charge (SoC) to ESS's charging and discharging powers. Limits on ESS charging ($P_t^{ESS,ch}$) and discharging ($P_t^{ESS,disch}$) rates are handled by equations (4) and (5). Moreover, limits on ESS's SoC are handled by equations (8) and (9). In addition, equation (7) ensures ESS has sufficient energy for the next day operation.

$$P_t^{ESS,used} + P_t^{ESS,sold} = P_t^{ESS,disch} DE_{ESS} , \forall t \quad (3)$$

$$P_t^{ESS,ch} \leq CR_{ESS} u_t^{ESS} , \forall t \quad (4)$$

$$P_t^{ESS,dis} \leq DR_{ESS} (1 - u_t^{ESS}) , \forall t \quad (5)$$

$$SOE_t^{ESS} = SOE_{t-1}^{ESS} + P_t^{ESS,ch} CE_{ESS} \Delta t - P_t^{ESS,dis} \Delta t \quad \forall t \quad (6)$$

$$SOE_t^{ESS} = SOE_t^{ESS,ini} , if\ t = 1 \quad (7)$$

$$SOE_t^{ESS} \leq SOE^{ESS,\max} , \forall t \quad (8)$$

$$SOE_t^{ESS} \geq SOE^{ESS,\min} , \forall t \quad (9)$$

where $DE_{ESS}$ and $CE_{ESS}$ are the efficiency of ESS charging and discharging cycles. $CR_{ESS}$ and $DR_{ESS}$ are ESS charging and discharging rates. $u_t^{ESS}$ is a binary variable determines charging and discharging status. $SOE_t^{ESS}$ is the ESS SoC that represent the amount of energy in ESS. $SOE^{ESS,\min}$ and $SOE^{ESS,\max}$ are the minimum and maximum energy that can be stored in ESS. $SOE_t^{ESS,ini}$ is the initial energy in the battery.

### D. EV Model

The proposed model for EV is pretty much similar to ESS model with the exception of specifying a time window [8:00 PM-8:00AM] for charging and discharging. EV must be fully charged when departures the home and it has 80 percent charge when it arrives.

$$P_t^{EV,used} + P_t^{EV,sold} = P_t^{EV,disch} DE_{EV} \quad \forall t \in [T^a, T^d] \quad (10)$$

$$P_t^{EV,ch} \leq CR_{EV} u_t^{EV} , \forall t \in [T^a, T^d] \quad (11)$$

$$P_t^{EV,dis} \leq DR_{EV}(1-u_t^{EV}) \ , \forall t \in [T^a, T^d] \tag{12}$$

$$SOE_t^{EV} = SOE_{t-1}^{EV} + P_t^{EV,ch} CE_{EV} \Delta t - P_t^{EV,dis} \Delta t \ , \forall t \in [T^a, T^d] \tag{13}$$

$$SOE_t^{EV} \leq SOE^{EV,\max} \ , \forall t \in [T^a, T^d] \tag{14}$$

$$SOE_t^{EV} \geq SOE^{EV,\min} \ , \forall t \in [T^a, T^d] \tag{15}$$

$$SOE_t^{EV} = SOE_t^{EV,ini} \ , if \ t = T^a \tag{16}$$

$$SOE_t^{EV} = P_t^{EV,used} = P_t^{EV,sold} = P_t^{EEV,disch} = P_t^{EEV,ch} = 0, \forall t \notin [T^a, T^d] \tag{17}$$

Similar to ESS model, two different operational modes including EV2Grid, EV2Home are considered. Accordingly EV discharged power can be used to electrify loads at home ($P_t^{EV,used}$) or inject to grid ($P_t^{EV,sold}$). $P_t^{EV,disch}$ and $DE_{EV}$ are EV's discharging power and discharging efficiency. Similarly, $P_t^{EV,ch}$ and $CE_{EV}$ are EV's charging power and charging efficiency. $CR_{EV}$ and $DR_{EV}$ are charging and discharging rates for EV. $u_t^{EV}$ is a binary variable determines the charging and discharging status of EV. $SoE_t^{EV}$ is the state of energy which represents the amount of energy in EV. $SOE_t^{EV,ini}$ is the initial energy in the EV. $SOE^{EV,\min}$ and $SOE^{EV,\max}$ are the minimum and maximum energy that can be stored in the EV. $T^a$ and $T^d$ are the arrival and departure time of the EV, respectively.

### E. PV Model

The proposed PV model is based on the energy conservation principle that balances the power at each time interval. In other words, energy generated by solar panels should be equal to total withdrawal energy from solar panels.

$$P_t^{PV,used} + P_t^{PV,sold} = P_t^{PV,gen} \ , \forall t \tag{18}$$

Similar to ESS and EV powers the generated power by solar panels can be used to electrify loads at home ($P_t^{PV,used}$) or inject to grid ($P_t^{PV,sold}$). $P_t^{PV,gen}$ is the power produce by the solar panels at $t^{th}$ time interval.

### F. Total Power Injected to Grid

Total injected power to grid should be equal to summation of $P_t^{PV,sold}$, $P_t^{ESS,sold}$, and $P_t^{EV,sold}$.

$$P_t^{sold} = P_t^{PV,sold} + P_t^{ESS,sold} + P_t^{EV,sold} \ , \forall t \tag{19}$$

where $P_t^{sold}$ is the total power inject to grid at $t^{th}$ time interval.

### G. Power Transaction Restriction

Withdrawing and injecting electricity from or to grid can not be done at the same time which is handled by equations (20) and (21).

$$P_t^{grid} \leq N_1 u_t^{grid} \ , \forall t \tag{20}$$

$$P_t^{sold} \leq N_2 (1 - u_t^{grid}) \ , \forall t \tag{21}$$

where $P_t^{grid}$ is the withdrawal power from the grid, $u_t^{grid}$ is a binary variable determines the status of withdrawing or injecting power from or to the grid. $N_1$ and $N_2$ are constants that restrict the withdrawal and injected power.

### H. Load Shifting Model

The proposed model for load shifting is based on the assumption that the energy requirement of the load may be shifted to a different time period within the acceptable time horizon. The load energy requirement is not reduced by deferral; it is merely delayed. Equations (22)-(28) model represent the proposed model for load shifting.

$$Pt^L = P_t^{ND} + PD_t^D + P_t^S \ , \forall t \in T \tag{22}$$

$$P_t^D = \sum_a P_t^{D,a} \ , \forall t \in T \tag{23}$$

$$P_t^S = \sum_{a \in A} \sum_{t'=t-ADT_a}^{t-1} P_{t',t}^{D,a} - \sum_{a \in A} \sum_{t'=t+1}^{t+ADT_a} P_{t,t'}^{D,a} \ , \forall t \in T \tag{24}$$

$$0 \leq \sum_{t'=t+1}^{t+ADT_a} P_{t,t'}^{D,a} \leq P_t^{D,a} \ \ \forall a \in A, t \in T \tag{25}$$

$$P_{t,t'}^a = 0, \ \forall a \in A, t' < t \tag{26}$$

$$P_{t,t'}^a = u_{t,t'}^a \times PD_t^a \tag{27}$$

$$\sum_t u_{t,t'}^a = 1 \ , t' \leq t + ADT \tag{28}$$

where, $a$ and $A$ are the index and set of responsive appliances, respectively. $PD^t$ is composed of $PD_t^{ND}$ (non-deferrable load), $PD_t^D$ (normally scheduled deferrable load), and $PS_t^S$ (load that has been shifted into or out of interval t). Based on this definition $PS_t^S$ can take both positive and negative values. The (constant) $P_t^D$ is the sum of the individual appliance loads $P_t^{D,a}$ in the set A of all deferrable appliances normally scheduled at time t. $P_t^S$ is the sum of all loads shifted into time t from previous times $t'$ less the sum of flexible loads shifted to future times. where $P_{t_1,t_2}^{D,a}$ is the load of appliance a that is shifted from $t_1$ to $t_2$ and $ADT_a$ is the acceptable delay time of appliance a. The first summation on the right-hand side of Equation (24) represents the normally scheduled load that has been shifted from previous time intervals to the current time interval, whereas the second summation represents the normally scheduled load that has been shifted from the current time interval to future time intervals. Equation (25) ensures the shifted load cannot exceed the normally scheduled deferrable load. In addition applying Equation (26) guarantee that load can only be delayed and should not be shifted to an earlier time interval .

## III. NUMERICAL RESULTS

Four different appliances including dishwashers (4), clothes washers (3), clothes dryers (1.5), and heating and ventilation systems (1.5) are considered as deferrable loads to validated the proposed approach for home energy management problem. Numbers in parenthesis specified the ADT values for appliances. To implement the DSM, a load profile is specified for each appliance which can be obtained in [19].

The electricity price in different time intervals are shown in (figure 1}). Note that price for selling electricity to the grid is equal to 3 ¢ for all time intervals. General Algebraic Modeling System (GAMS), a high-level modeling system for mathematical programming and optimization is used to solve the proposed mixed integer linear programming optimization problem. Different combination of EV, ESS and PV are considered in several cases studies to show the individual effect of PV, EV and ESS on the customer's electricity cost.

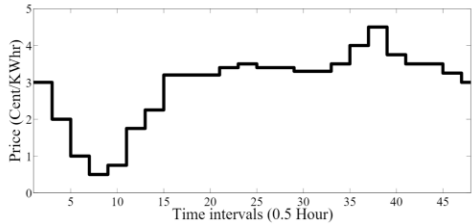

Figure 1. Electricity price at different time intervals.

### A. Base Case

Neither PV nor storage systems (including EV and ESS) are considered in this case. In this case demand side management is the only factor that plays a role on reducing electricity cost. Electricity bill can be reduced by shifting loads from the expensive time interval to cheaper time intervals. The electricity cost before applying DSM is 369.335 ¢ and it reduces to 364.835 ¢ after implementing DSM. As it was expected electricity cost is not reduced significantly in this case because ADT-related constraints restrict load shifting. Load profile before and after applying DSM is shown in figure 2. It is clear that after implementing DSM some loads have been shifted from expensive time intervals to cheaper ones.

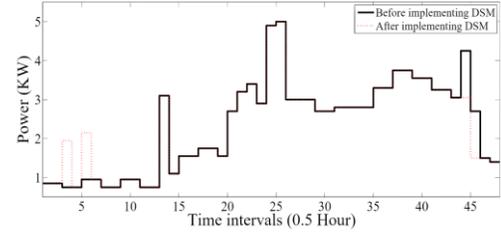

Figure 2. Electricity price at different time intervals.

### B. Effect of PV Power on Electricity Cost

A solar electric system has a uniquely quantifiable benefit: electricity that powers the home. A solar electric system can significantly reduce the cost of ownership. The effect of solar panels on the electricity cost is studied in this section. The electricity cost after considering PV is dropped to 90.075 ¢ and it drops more to 81.157 ¢ while the DSM effect is considered as well. It is obvious that the effect of PV on electricity cost is significant. Another important thing is the effect of DSM is more tangible in this case (i.e., it reduces the electricity cost by 8.918¢ ). The different power including purchased and sold power from and to grid are shown in figure 3 and figure 4 for the cases with and without DSM. From figure 3 it is clear that PV could satisfy loads in mid-intervals of day, which are the most expensive intervals, and that is why the electricity cost is dropped significantly. Another reason for dropping electricity cost by PV is selling the surplus of the photovoltaic power to the grid which is depicted by the green color in figure 3.

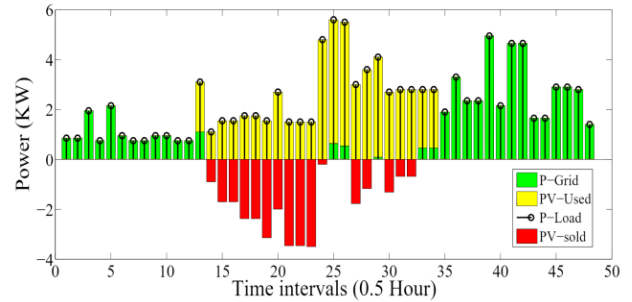

Figure 3. Optimal scheduled powers in Case B with implementing DSM.

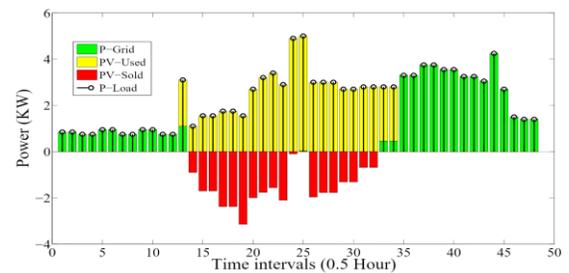

Figure 4. Optimal scheduled powers in Case B without implementing DSM.

Augmenting PV panels in home can decrease electricity cost by 77.75 %, which can be considered as a deep plunge in electricity cost. Thus a considerable amount of money can be saved by installing PV panels. The Return-On-Investment (ROI) is a factor that measures how effectively the firm uses its capital to generate profit; the higher the ROI, the better. The RIO for PV systems is high since it can return the money has been invested to install them by reducing electricity cost.

Comparing figure3 and figure 4 revels that loads at expensive time intervals are shifted to cheaper time intervals that can prevent purchasing power with more expensive prices.

## C. Effect of PV and ESS on electricity cost

ESSs can make usage of photovoltaic electricity generation more profitable by making this generation coincident with the peak load demand. Furthermore, arbitrage using augmenting ESS and PV that is done by storing photovoltaic generation in ESS and sell it back to grid on time intervals with expensive electricity price, can make implementing ESS and PV panels even more exploitable. Thus, investigating the impact of PV panels and ESS implementation is become more interesting. Note that the effect of EVs is discarded in this case since a multitude of customers still do not have an EV. Furthermore, investigating the impact of photovoltaic generation incorporated with ESS on electricity cost is an absolute necessity since PV panels usually install with ESS as a package.

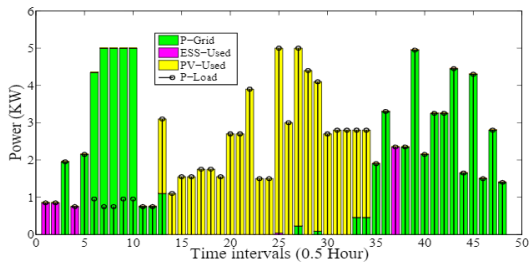

Figure 5. Optimal scheduled powers in Case C with implementing DSM.

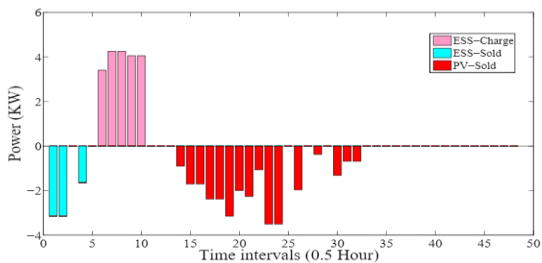

Figure 6. Optimal scheduled powers in Case C with implementing DSM.

Figure 5 and Figure 6 illustrate the optimum scheduled power after application of DSM. As it was expected at the middle of day, when sunlight is high, almost all the power demand at home is provided by PV panels. Furthermore, the surplus of PV power sells back to grid to make more profit. Like other case studies, ESS gets charge and discharge to make profit through tariff arbitrage. Electricity costs after applying DSM is equal to 53.309 ¢ while discarding the DSM impact increases the electricity cost to 62.55 ¢.

Comparing electricity cost with and without considering DSM impact reveals the importance of incorporating DSM. Note that reduction in electricity cost is obtained by rescheduling few appliances' demand. Increasing the size of ESS can reduce electricity cost at the cost of more capital investment. There is also a cap for ESS size at home in which further increase in ESS size would not reduce the electricity cost [20-24]. The optimum size of ESS depends on different factors such as peak demand, PV panel size, etc.

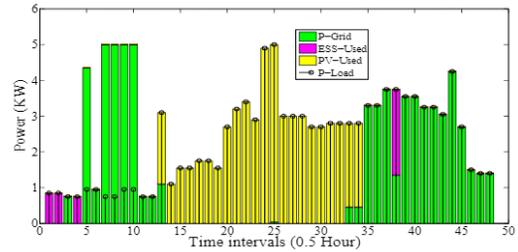

Figure 7. Optimal scheduled powers in Case C without implementing DSM.

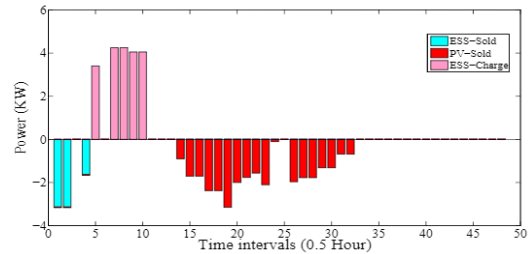

Figure 8. Optimal scheduled powers in Case C without implementing DSM.

Augmenting ESS with PV panels increases PV panels' profitability by storing photovoltaic energy to use in expensive time intervals. For Optimal scheduled power without considering DSM are shown in Figure 7 and Figure 8. Comparing these figures with Figure 5 and 6 demonstrates the impact of DSM on load profile. DSM is able to shave peak demand at noon to facilitate more injecting power to grid at expensive time intervals result in reducing electricity cost.

## D. Effect of PV, ESS and EV on electricity cost

The overall energy storage capacity increases by augmenting EV in an ESS equipped home. Thus, DSM program has more capacity for reshaping load profile to reduce electricity cost. The optimal scheduled power after applying DSM are depicted in Figure 9 and Figure 10 for the case with DSM. EV and ESS charge during the cheap intervals (i.e., the early hours) and use their stored electricity in more expensive time intervals either for electrify a portion of load at home or sell it back to the grid. The total injected power to grid with and without applying DSM are 38.364 KW and 36.42 KW, respectively. Compare to case C it is clear that installing ESS and EV can pave the ground for injecting more power to the grid. In other words augmenting energy storage systems facilitates customers to fully exploit PV panels.

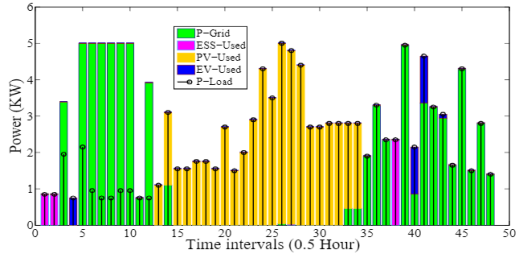

Figure 9. Optimal scheduled powers in Case D with implementing DSM

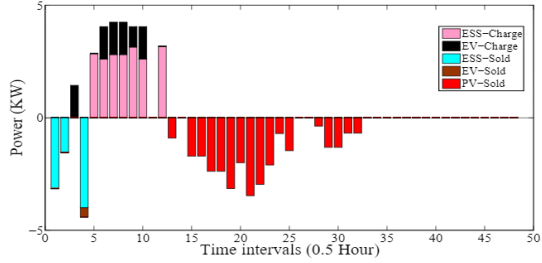

Figure 10. Optimal scheduled powers in Case D with implementing DSM.

The optimal scheduled power excluding DSM impacts are shown in Figure 11 and Figure 12. Comparing these figures with Figure 9 and Figure 10 indicates that application of DSM facilitate EV to inject power to grid. Yet another important point is DSM application increases the power provided by EV to be used at home by to house by 28% that implies further exploitation of EV capacity. Further exploitation of ESS capacity can be reached by incorporating DSM program. Thus, customers who has invested on ESS and EV can make more money by better exploiting these devices through implementing DSM management programs. Furthermore, DSM increase the return on investment rate for ESS and EV that can be encouraging for customers to invest on ESS and EV.

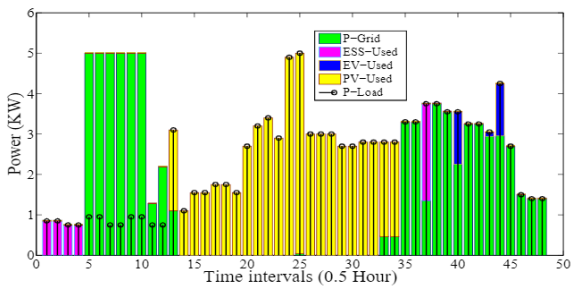

Figure 11. Optimal scheduled powers in Case D without implementing DSM.

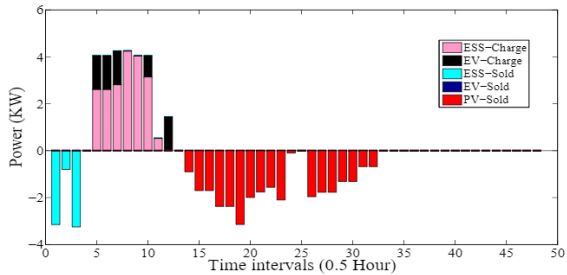

Figure 12. Optimal scheduled powers in Case D without implementing DSM.

## IV. CONCLUSION

A comprehensive mixed integer linear programming based approach for demand side management at home incorporated with PV panels, EV, and ESS is introduced to minimize electricity cost. Bi-directional power injection capability is implemented for both EV and ESS devices that makes the proposed approach more practicable. A detailed model based on acceptable deferral time for each appliances is presented to take customers' comfort into account. The proposed model is flexible to customers' preference to determine new acceptable deferral time for different appliances. These characteristics distinguish the proposed approach from those in literature and make it more practicable.

Several comparisons underscore the contributions of EV, ESS, PV, and DSM program on electricity cost. It has been shown that the largest contributor to the electricity cost reduction is PV panel in which can reduce the electricity cost reduces by 25% relative to the base case. The impacts of EV and ESS on electricity cost reduction are not comparble with PV panel's impact. However, incorporating these devices with PV panels makes them impactful.

Ongoing work aims to consider probable uncertainties including uncertainties in photovoltaic energy, and departure and arrival time of EV.